\documentclass[global,twocolumn]{svjour}
\usepackage{graphics}
\usepackage{amsmath}
\usepackage{color}
\usepackage{subscript}
\begin{document}
\title{Phase-dependent spectral control of pulsed modulation instability via dichromatic seed fields}
\author{Maximilian Brinkmann \and Michael Kues \and Carsten Fallnich
}                     
%
%
\institute{Institute of Applied Physics, Westf\"alische Wilhelms-Universit\"at M\"unster, Corrensstra\ss e 2, 48149 M\"unster, Germany}
\date{}
\maketitle

\begin{abstract}
We investigated experimentally and numerically the spectral control of modulation instability (MI) dynamics via the initial phase relation of two weak seed fields.
Specifically, we show how second-order modulation instability dynamics exhibit phase-dependent anti-correlated growth rates of adjacent spectral sidebands. This effect enables a novel method to control MI-based frequency conversion: in contrast to first-order MI dynamics, which exhibit a uniform phase dependence of the growth rates, second-order MI dynamics allow to redistribute the spectral energy, leading to an asymmetric spectrum. Therefore, the presented findings should be very attractive to different applications, such as phase-sensitive amplification or supercontinuum generation initiated by MI.
\end{abstract}

\section{Introduction}
Modulation instability (MI) \cite{Zakharov2009}, describing the exponential amplification of a weak modulation on a strong pump field background in a nonlinear dispersive medium, is spectrally characterized by a drastic energy transfer from the pump's carrier frequency $\nu_0$ into spectral sidebands around $\nu_0$. Experimentally this process can be induced by superimposing the pump field with a weak seed field, frequency-shifted relative to $\nu_0$, to imprint a defined initial modulation on the pump's amplitude. In this way, initiated by a weak coherent signal, MI is exploited, e.g. for ultrashort pulse train generation \cite{Tai1986a} or in seeded supercontinuum generation \cite{Dudley2008}.

When seeding with a single-frequency mode, the MI evolution can be controlled via the amplitude of the seed field and via the frequency shift of the seed field, defining the spectral positions of the sidebands. As pointed out in reference \cite{Erkintalo2011a} the MI evolution can furthermore be spectrally controlled via the pump-seed phase relation, when excited by two seed frequencies symmetrically spaced around $\nu_0$. Specifically, a first-order MI excited by two symmetrically placed seed frequencies exhibits symmetric sideband-growth depending on the pump-seed phase relation, providing a possibility to change the amplitudes of all sidebands in equal measure.

In contrast to the conditions stated above, a second-order MI is excited, if two seed frequencies do not lie symmetrically around the carrier frequency, but are spectrally located together either on the low or high frequency side of the carrier frequency \cite{Erkintalo2011}. Here, we show numerically and experimentally that second-order MI dynamics exhibit an antisymmetric phase dependence of the growth rates, meaning that the growth rates of two sidebands, which form a sideband pair, vary anti-correlated with the initial phase. This presented effect enables a novel method to control MI-based frequency conversion and should therefore be attractive to different applications.

The nonlinear element required for the excitation of such MI dynamics was chosen to be a microstructured fiber (NL-PM-750, NKT-Photonics). Furthermore, in order to circumvent high continuous-wave powers we used pulsed pump and seed fields to excite MI dynamics, which is a common method especially in investigations on MI-based supercontinuum generation \cite{Solli2010a}. Note that MI dynamics ultimately lead to a fission of the pump pulse after a certain propagation distance into individual soliton-like subpulses due to disruption by higher-order dispersion and Raman scattering \cite{Dudley2006}. After the fission of the pulse the optical field dynamics are not governed by MI anymore, thus we limited our investigations to the dynamics before the inset of pulse fission.

This paper is structured as follows: first, the applied numerical model is outlined in section \ref{sec:num} and the used experimental setup is presented in section~\ref{sec:setup}, subsequently, the uniform phase dependence of the growth rates of first-order MI are experimentally verified in section~\ref{sec:firstmi} and the investigation on second-order like MI is presented in section~\ref{sec:secondmi}.

\section{Numerical model}
\label{sec:num}
In order to model and numerically investigate MI dynamics in a microstructured fiber we used the generalized scalar nonlinear Schr\"odinger equation (GNLSE), which has been proven many times to accurately describe nonlinear unidirectional pulse propagation in MSF \cite{Dudley2006}. Explicitly including higher-order linear and nonlinear terms, we used the GNLSE in the following form \cite{Dudley2006}:
\begin{align}
	\label{GNLSE}
	 \nonumber 	 \frac{\partial A}{\partial z}=&-\frac{\alpha}{2}A+\sum_{k\geq2}^{10}\frac{i^{k+1}}{k!}\beta_k\frac{\partial^k A}{\partial t^k}
	 +i\gamma\left(1+i\frac{1}{\nu_0}\frac{\partial}{\partial t}\right)\\
	 &\cdot\left(A(z,t)\int_{-\infty}^{\infty}R(t')\left|A(z,t - t')\right|^2dt'\right).
\end{align}
Considering the slowly varying envelope approximation, here, $A(z,t)$ describes the pulse envelope and $t$ is the time in a frame of reference moving with the group velocity $\frac{1}{\beta_1}$. The values $\beta_k$ are the dispersion coefficients at the center frequency $\nu_0$ and $\alpha$ and $\gamma$ are the absorption and nonlinear coefficients of the fiber. The response function $R(t)=(1-f_R)\delta(t)+f_R h_R(t)$ with $f_R=0.18$ includes both instantaneous electronic and delayed Raman contributions, whereby we used the analytic form of the Raman response function \cite{Blow1989}: $h_R(t)=[(\tau_1^2+\tau_2^2)/(\tau_1\tau_2^2)]\exp(-t/\tau_2)\sin(t/\tau_1)$ with $\tau_1=12.2$\,fs and $\tau_2=32$\,fs.

We solved the GNLSE numerically stepwise along the fiber by means of a split-step Fourier method \cite{Agrawal-NL-Fibre-Optics}, considering a step size of $10\,\mu$m, an array size of $2^{16}$ points and a temporal resolution of 0.5\,fs.
The modeled fiber input fields constituted superpositions of $\text{sech}^2$-shaped pulses, representing the pump and seed pulses. In order to investigate the field evolution depending on the relative phase between the pump and the seed pulses, the phase of the pump pulse was varied by multiplying time-independent phase terms to the initial pump pulse, while keeping the phase of the seed fields constant.
Thus, the following investigations were concentrated only on the effect of the relative phase between the pump and both seed pulses, whereby the relative phase between the two seed pulses was not varied. Indeed, by means of our numerical model we were able to verify that the field evolution is determined by the phases of all three interacting pulses relative to each other. However, varying the phase of the pump pulse accomplished the simplest and most effective spectral control of the MI dynamics. Therefore, we restricted our investigations to control the individual sidebands in a correlated or anti-correlated way on the relative phase of the pump pulse.

\section{Experimental setup}
\label{sec:setup}
\begin{figure*}[htb]
\centerline{\includegraphics{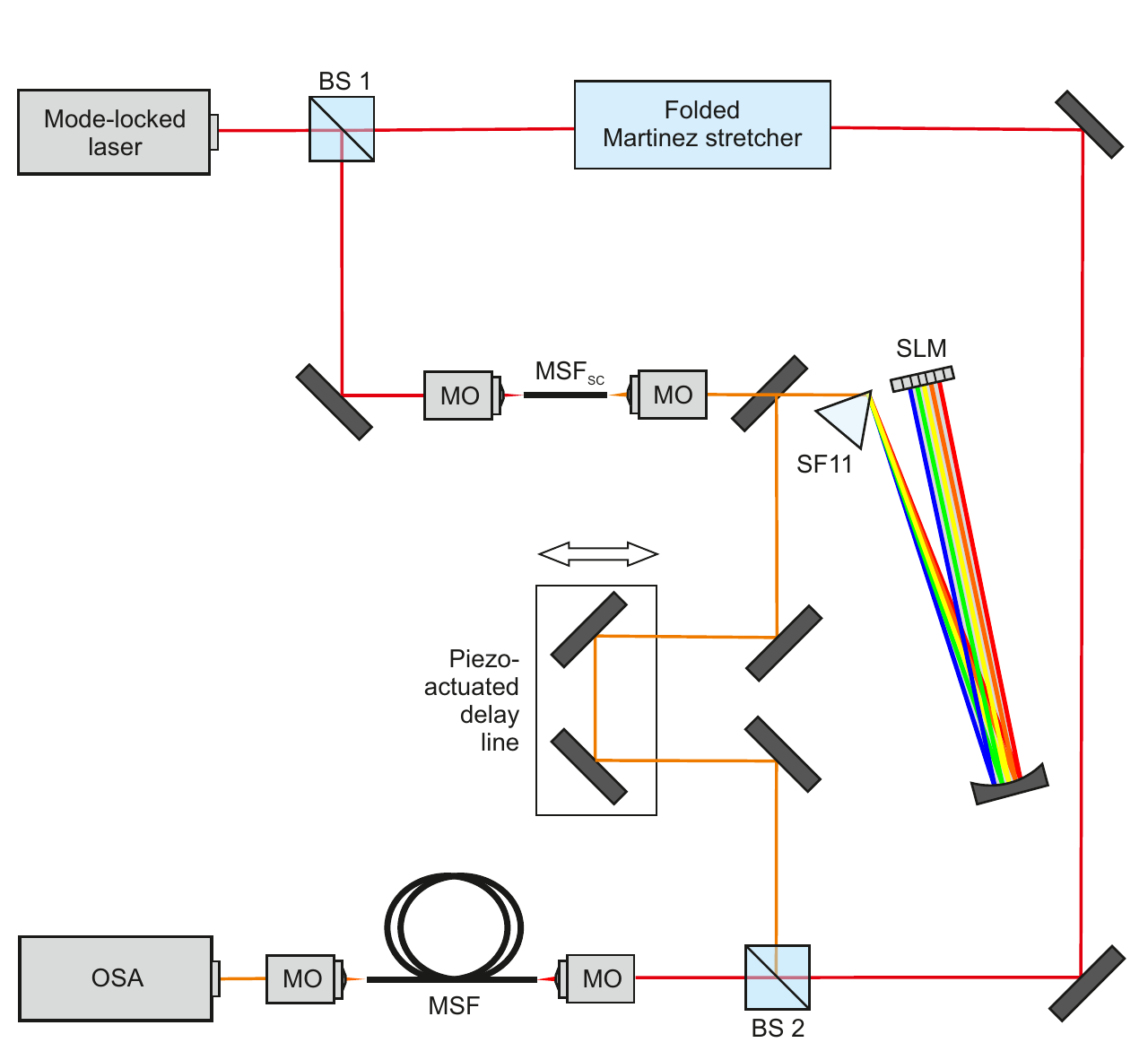}}
\caption{\label{fig:exp_setup}Schematic diagram of experimental setup. BS: beam splitter, MO: microscope objective, MSF: microstructured fiber used to investigate MI dynamics, MSF\textsubscript{SC}: microstructured fiber used to generate supercontinuum pulses, SLM: spatial light modulator, OSA: optical spectrum analyzer.}
\end{figure*}
A schematic of the experimental setup is shown in Fig.~\ref{fig:exp_setup}: the microstructured fiber (NL-PM-750, NKT-Photonics) used to investigate MI dynamics had a length of $25$\,cm, the nonlinear coefficient was $\gamma=0.095\,\text{/(Wm)}$, and the second-order dispersion was $\beta_2=-0.5812\,\text{ps}^2\text{/cm}$ at $384.5$\,THz, i.e. at a wavelength of $782$\,nm.
To allow for mutually coherent pump and seed fields a mode-locked titanium-sapphire laser was used, emitting bandwidth-limited pulses with a duration of $80$\,fs (FWHM) at a center frequency of $384.5$\,THz. 
The pump pulses were generated by stretching the laser output pulses to a temporal duration of $1.8$\,ps, necessary to suppress self-phase modulation within the fiber so that a field evolution dominated by MI was ensured. The pulse stretching was accomplished by employing normal dispersion via a folded Martinez stretcher \cite{martinez}, consisting of a cylindrical lens ($f=100$\,mm) and a transmission grating (1200 lines per \,mm). In this way, the fiber was pumped with a chirped pulse in the anomalous dispersion regime. However, we spend no further attention to the pulse chirp as we were able to verify by numerical investigations that a pulse chirp does not alter the phase dependence of the MI evolution. 

To generate the seed fields, replicas of the laser output pulses produced with a beam splitter (BS~1) were focused into an additional segment (MSF\textsubscript{SC}, length of 3\,cm) of the mentioned microstructured fiber to allow for coherent supercontinuum generation by exploiting soliton dynamics. The seed fields were then synthesized by cutting the desired frequencies out of the supercontinuum spectrum using a spectral pulse shaper based on a SF11 prism and a liquid crystal spatial light modulator (SLM). 
Before the pump field and the seed fields were combined and simultaneously injected via a 40x microscope objective into the microstructured fiber to excite MI dynamics, the pump field passed a piezo-actuated retroreflector enabling phase changes between the pump and seed fields. The spectra of the fiber output pulses were measured with an optical spectrum analyzer.

\section{First-order modulation instability}
\label{sec:firstmi}
\begin{figure}[htb]
\centerline{\includegraphics{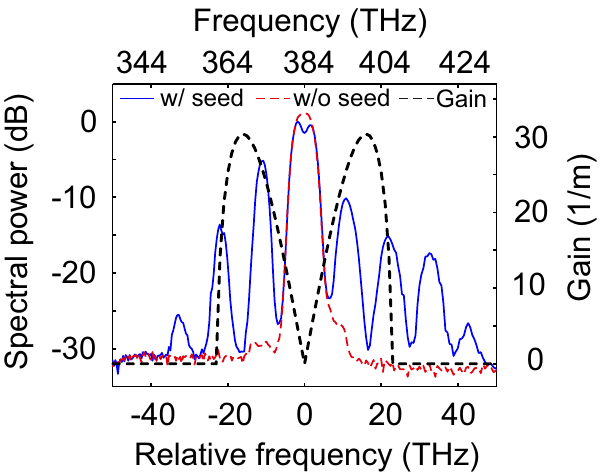}}
\caption{\label{fig:MI1s}Measured spectra without (red-dashed line) and with seed at $\nu_{-1}=-10.5$\,THz (blue solid line), as well as MI gain curve for the used fiber and pump pulse peak power (black dashed line).}
\end{figure}

Figure~\ref{fig:MI1s} shows a measured unseeded fiber output spectrum (red dashed line) for an estimated pump peak power of $P_0=300$\,W, and the associated MI gain \cite{Dudley2006} (black dashed line). In contrast to this unseeded case, the measured spectrum illustrated by the blue-solid line was generated by the pump field superimposed with a single seed field, whose center frequency of 374.0\,THz was located within the gain bandwidth. The average seed power was about 1000 times smaller than the average pump power of about 45\,mW. The single seed field stimulated a cascaded generation of sidebands around the pump frequency. 
Specifically, the spectrum of the seeded case shows three pairs of sidebands, which are spectrally separated by the induced modulation frequency $\Omega=10.5$\,THz, clearly exposing a successful excitation of pulsed MI.
However, in contrast to the commonly known symmetric MI spectra \cite{Dudley2009} generated by continuous pump and seed fields, the spectrum in Fig.~\ref{fig:MI1s} shows an additional sideband standing alone on the high frequency side of the pump frequency mode at about 42\,THz without a partner on the low frequency side. We were able to attribute this symmetry breaking by means of numerical simulations to effects of higher-order dispersion and Raman scattering.
Nevertheless, the pump frequency mode in the spectrum of the seeded case is furthermore depleted compared to the spectrum of the unseeded case, which is another characteristic of MI owing to the energy transfer into the sidebands. In the following, these sidebands shall be denoted relative to $\nu_0$ by their frequency shift: $\nu_n=n \cdot \Omega$ with $n=\pm1,\pm2,\pm3$, ... .

\begin{figure}[htb]
\centerline{\includegraphics{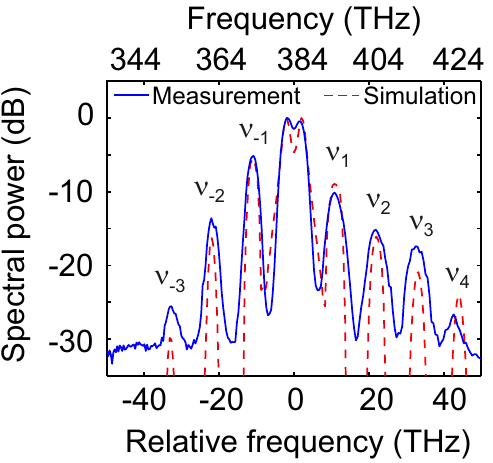}}
\caption{\label{fig:vgl}Measured (blue solid line) and simulated (red dashed line) spectra with seed at $\nu_{-1}=-10.5$\,THz.}
\end{figure}

The experimentally induced MI dynamics could also be modeled accurately with our numerical model: considering the above presented experimental parameters, simulations yielded the dash-drawn spectrum in Fig.~\ref{fig:vgl}. Besides the measured and simulated spectra, Fig.~\ref{fig:vgl} furthermore illustrates the notation of the sidebands. It can be seen that the amplitude levels of the measured and simulated spectra do not coincide accurately, which can be accounted for by inaccuracies in the experimental parameters: especially the seed field's temporal duration and peak power could only be approximated because of the low peak power. However, as the measured spectral position of the sidebands are well reproduced by the simulations, and as we are only interested in relative amplitude changes, the simulations constitute an adequate tool for estimation and comparison.

\begin{figure}[htb]
\centerline{\includegraphics{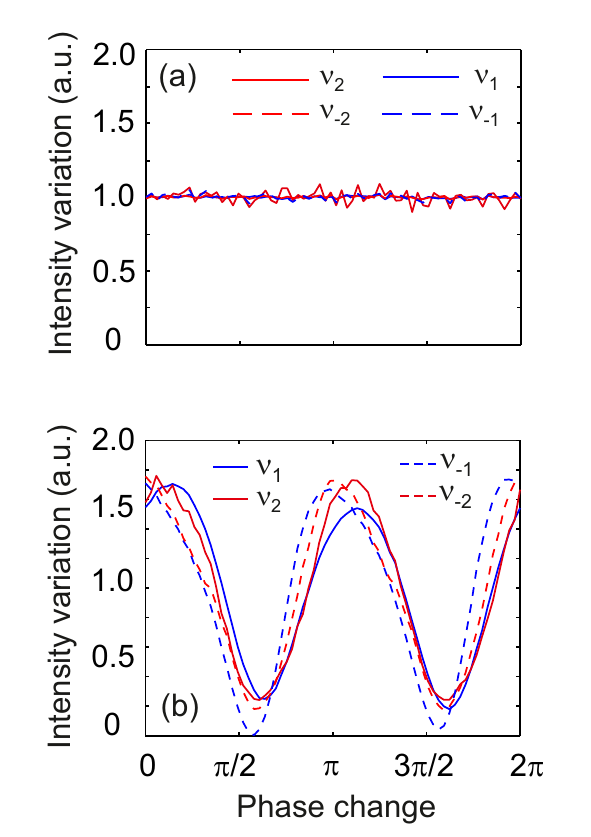}}
\caption{\label{fig:sim_2ssym} Simulated intensity variation of sidebands $\nu_{\pm 1}, \nu_{\pm2}$, seeded with (a) one seed field at $\nu_{-1}=-10.5$\,THz  and (b) two seed fields at $\nu_{-1}=-10.5$\,THz and $\nu_{1}=10.5$\,THz.}
\end{figure}

Based on this agreement, in a first step numerical investigations focusing on a spectral phase dependence were performed, considering a pump peak power of $P_0=300$\,W (average power of $\bar{P}_0=45$\,mW) and a single seed at $\nu_{-1}=-10.5$\,THz with an average power of $\bar{P}_{\text{Seed}}=\frac{1}{1000}\bar{P}_0$. Note, that only the average power of the seed pulse is stated as the peak power of the pulse could only be estimated.
However, considering a spectral width of the seed pulse of 8\,nm (corresponding to a bandwidth-limited duration of 80\,fs), an upper limit of 7\,W for the seed peak power can be specified.
For this case, Fig.~\ref{fig:sim_2ssym} (a) shows the intensities of the first two sideband pairs as a function of the phase of the pump. All sideband intensities remain constant; thus a phase dependence is not observable. In contrast, Fig.~\ref{fig:sim_2ssym}~(b) contains the same diagram but for two seed fields at frequencies of $\nu_{-1}=-10.5$\,THz and $\nu_{1}=10.5$\,THz; both are lying under the MI gain curve and are fulfilling the frequency relation $\nu_{-1} = -\nu_{1}$. Here, the sideband intensities vary synchronously, attesting phase-dependent but positively correlated growth rates of the sidebands. We were able to attribute the slight phase-shifts between the four displayed curves by numerical simulations (not shown here) to higher-order dispersion and Raman effects.

\begin{figure}[htb]
\centerline{\includegraphics{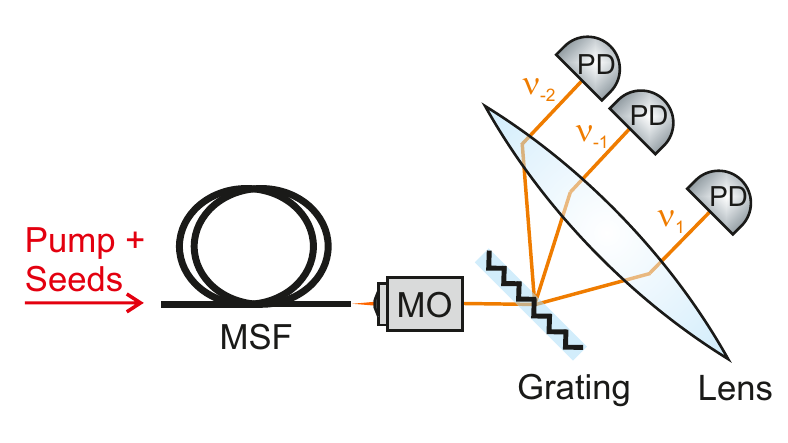}}
\caption{\label{fig:exp_setup_phase} Setup to measure phase-dependent variations of the intensities of the first three sideband modes ($\nu_{\pm 1}$, $\nu_{-2}$) of the fiber output spectrum. MO: microscope objective, MSF: microstructured fiber, PD: photodetector.}
\end{figure}

To allow for an appropriate experimental measurement of the variation of the sidebands' intensities as a function of the phase of the pump pulses, the slow optical spectrum analyzer was replaced with the setup illustrated in Fig.~\ref{fig:exp_setup_phase}: the spectral components of the fiber output were spatially dispersed employing a grating (1200\\lines~per~\,mm), and the sideband modes $\nu_{\pm 1}$ and $\nu_{-2}$ were individually focused onto separate fast silicon photodetectors (150\,MHz~bandwidth). All detectors were connected to an oscilloscope (1\,GHz~bandwidth). Due to the limited number of channels of this oscilloscope, only the intensities of the mentioned first three sidebands could be measured. 
The phase of the pump pulse was varied by modulating the piezo with a sinusoidal function, fast enough to neglect fluctuations of the interferometric setup caused by external mechanical or thermal perturbations over the measurement time.
In this way, phase-dependent effects could be investigated without an active stabilization of the setup.
Synchronously to the sidebands' intensities the piezo voltage was measured, allowing, in combination with a calibration of the delay induced by the piezo, for a reconstruction of the phase variation.

\begin{figure}[htb]
\centerline{\includegraphics{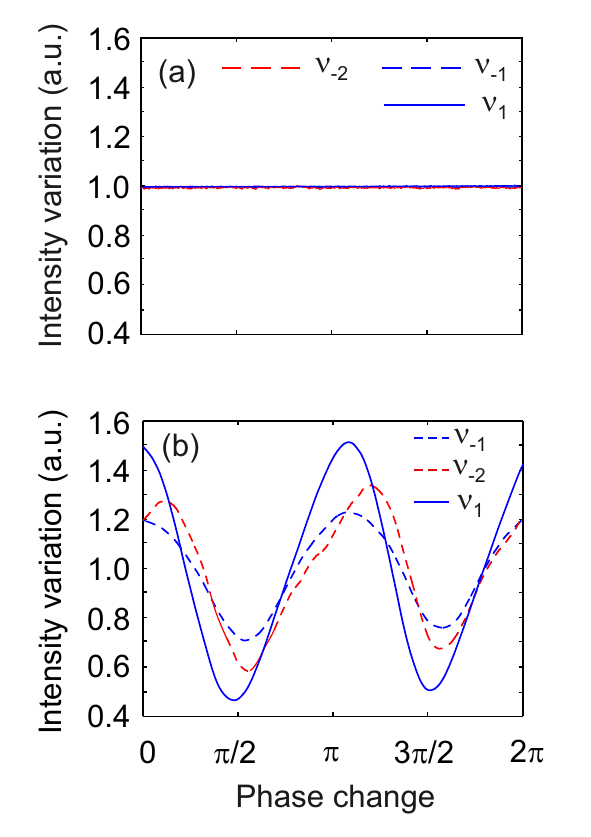}}
\caption{\label{fig:exp_2ssym} Measured intensity variation of sidebands $\nu_{\pm 1}, \nu_{-2}$, seeded with (a) one seed field at $\nu_{-1}=-10.5$\,THz  and (b) two seed fields at $\nu_{-1}=-10.5$\,THz and $\nu_{1}=10.5$\,THz , as a function of the phase shift of the seed fields relative to the pump.}
\end{figure}

For seeding the MI with only a single frequency mode at $\nu_{-1}=-10.5$\,THz and the same parameters as above, Fig.~\ref{fig:exp_2ssym}~(a) shows the detected signals of the sidebands normalized to their own mean value as a function of the phase variation. In accordance with the simulations, the intensities of the sidebands do not change. However, with the injection of a second seed at a frequency of $\nu_{1}=10.5$\,THz into the fiber a change of the sidebands' intensities becomes clearly visible as illustrated in Fig.~\ref{fig:exp_2ssym}~(b). As predicted by the numerical simulation the intensities of the sidebands do change in equal measure with the phase variation. The slight phase shifts between the displayed curves and the mismatch of the magnitude of variations of the three curves  were attributed to higher-order dispersion and Raman effects as in the numerical simulations.  
\newline

\section{Second-order modulation instability}
\label{sec:secondmi}
\begin{figure}[htb]
\centerline{\includegraphics{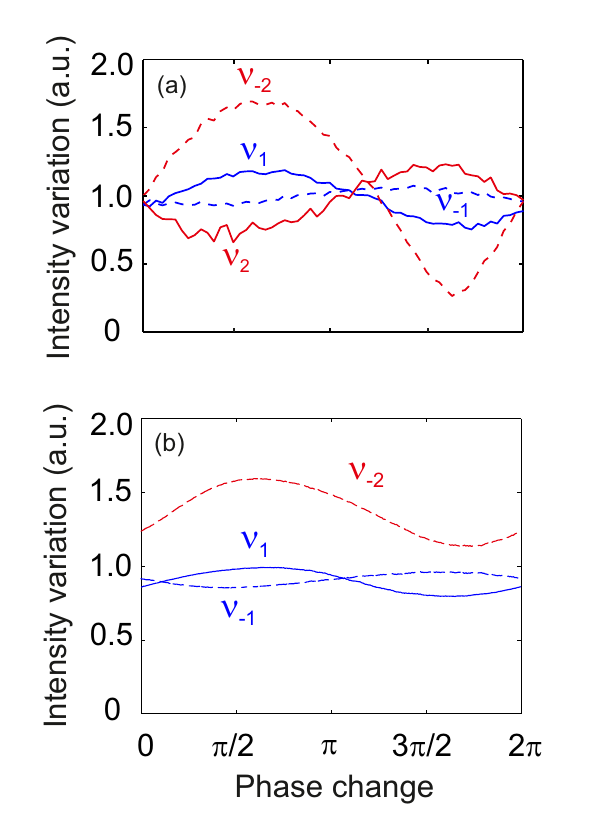}}
\caption{\label{fig:exp}Simulated (a) and measured (b) intensity variation of sidebands $\nu_{\pm 1}, \nu_{-2}$, seeded with two seed fields at $\nu_{-1}=-10.5$\,THz and $\nu_{-2}=-21.0$\,THz.}
\end{figure}

The uniform phase dependence of the growth rates of the sidebands changes by seeding with two frequency modes which are not symmetrically spaced around the pump frequency, but which lie on the same side of the pump. Considering a pump peak power of $P_0=370$\,W, and two seed fields at $\nu_{-1}=-10.5$\,THz and $\nu_{-2}=-21.0$\,THz, each with an average power of $\bar{P}_{\text{Seed}}=\frac{1}{1000}\bar{P_0}$ (with $\bar{P_0}$ denoting the average pump power of 58\,mW), simulations were executed. Note, that the second seed frequency is  a harmonic of the first ($\nu_{-1}=2\nu_{-2}$).
Figure~\ref{fig:exp}~(a) shows the revealed intensities of the first two sideband pairs as a function of the phase of the pump. To quantify the spectral change due to the second seed pulse, the intensity signals are normalized to the respective intensity signals obtained for the case of only one seed at $\nu_{-1}$ (not shown here, equal to Fig.~\ref{fig:exp_2ssym}~(a)).
In Fig.~\ref{fig:exp}~(a), revealed by our numerical model, the intensities change with the phase as well, but the intensities evolve contrary, i.e. whereas the intensity of $\nu_{1}$ increases the intensity of $\nu_{-1}$ decreases and vice versa (the same applies for $\nu_{2}$ and $\nu_{-2}$). Thus, the intensities of the two partners of a sideband pair (e.g. $\nu_1$ and $\nu_{-1}$) vary anti-symmetrically with the phase, resulting in an asymmetric spectral distribution. In the same way, adjacent sidebands (i.e. $\nu_{-1}$ and $\nu_{-2}$) also vary anti-symmetrically, which is in total contrast to first-order MI dynamics, in which the sidebands vary symmetrically.
Figure \ref{fig:exp} (b) shows the same diagram, but deduced from an experiment with the same parameters. As predicted by the simulation the intensities of the respective sideband pairs do evolve contrarily with the phase variation. The measured intensity of the first two sidebands changed by 20\% whereas the intensity of the third sideband changed by 40\% compared to the phase-independent signals obtained for the case of only one seed at $\nu_{-1}$. In contrast to the phase dependence of a first-order MI presented in section~3 (Fig.~\ref{fig:exp_2ssym}), the growth rates of adjacent sidebands are anti-correlated in this case.

A phase-dependent evolution was not observable, when choosing the second seed frequency to $\nu_{-3}=-31.5$\,THz, which was located outside of the  MI gain curve. In this case the MI dynamics are solely defined by the first seed frequency mode, showing again no phase dependence. This fact excludes the observed phase-dependent effect from being of interferometric character. Thus, the observed phase-dependent behavior is a clear evidence and a first experimental proof of phase-controlled second-order like MI dynamics.

It is worth to emphasize again the difference between the phase dependence of a first-order MI and of a second-order MI: as shown in Fig.~\ref{fig:exp_2ssym}~(b), a first-order MI, excitable by two seed frequencies ($\nu_{-1}$ and $\nu_{1}$) symmetrically spaced around the pump frequency, exhibits a uniform phase dependence of the sidebands. Consequently the spectral energy across the sidebands can be increased or decreased, but the spectral distribution cannot be changed with the phase, retaining the symmetry properties of the spectrum.
In a second-order MI, as shown in Fig.~\ref{fig:exp}, two sidebands, forming a pair, vary anti-symmetrically with the phase. Thus, the spectral distribution can be changed, leading to an asymmetric spectral distribution.

\begin{figure}[htb]
\centerline{\includegraphics{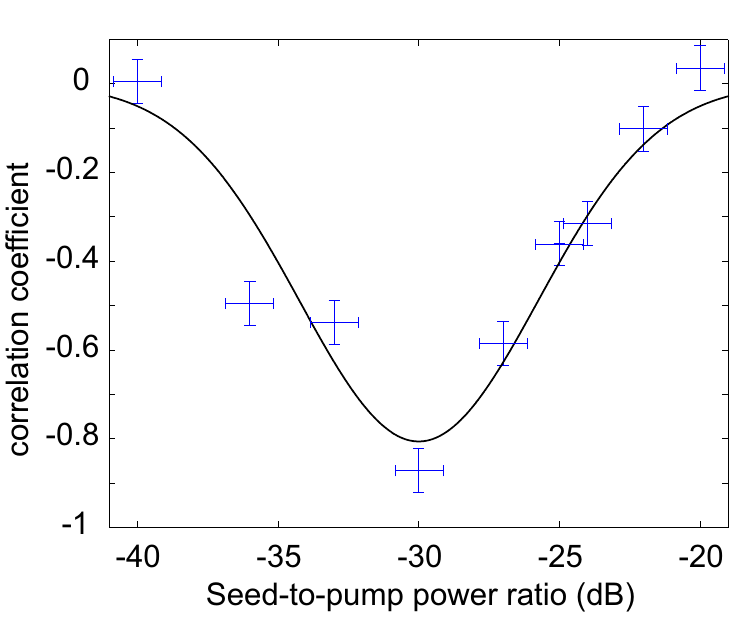}}
\caption{\label{fig:exp_seedpow} Correlation coefficient (blue crosses) of the measured intensities of the sidebands $\nu_{- 1}$ and $\nu_{-2}$ as a function of the seed power for a MI seeded with two seed fields at $\nu_{-1}=-10.5$\,THz and $\nu_{-2}=-21.0$\,THz. The black solid line is drawn to guide the eye.}
\end{figure}

Besides the spectral placement of the seed pulses, the power of the seed pulses was manifested as another crucial parameter to observe phase-dependent second-order MI dynamics.
Therefore, in order to enable the comparison of measured phase-dependent intensity variations for different seed powers, we quantified the antisymmetric variation of two adjacent sidebands by the correlation coefficient, defined as
\begin{equation}
	\text{corr}(I_n,I_m)=\frac{E\left[\left(I_n-\mu_n\right)\left(I_m-\mu_m\right)\right]}{\sigma_n\sigma_m}.
\end{equation}
Here, $E$ is the expected value operator, $\mu_j$ the mean value and $\sigma_j$ the standard deviation of the intensity $I_j$ of the sideband $\nu_{j}$, with $j=\pm1,\pm2, ...$ . The correlation coefficient can adopt a value in the interval $\left[-1;1\right]$, with a value of $1$ denoting perfect correlation and a value of $-1$ denoting perfect anticorrelation. 
Figure~\ref{fig:exp_seedpow} shows the correlation coefficient $\text{corr}(I_{-1},I_{-2})$ for the signals of $\nu_{-1}$ and $\nu_{-2}$ as a function of the seed power and the experimental parameters as above. With a correlation coefficient of -0.9, the investigated phase dependence was most distinct for an average seed power of about 1000 times smaller than the average pump power. The corresponding measured signals were shown in Fig.~\ref{fig:exp}.
Increasing the seed power above 1/1000 of the pump power degraded the correlation, which can be accounted for by the onset of soliton dynamics interfering with the MI evolution, as a higher seed power reduces the propagation length up to the point, at which the pulse splits into individual soliton-like sub-pulses. 
Also lowering the seed power below 1/1000 of the pump power led to a reduced correlation due to a lower seed-power-to-noise ratio, which degraded the influence of the seed field. Decreasing the seed-to-pump power ratio below 1/10000, suppressed the phase dependence of the sidebands. 
To illustrate the outlined tendency, caused by the competition between seeded MI, noise-driven MI and soliton dynamics, we fitted the solid curve to the data points in Fig.~\ref{fig:exp_seedpow}.

\section{Conclusion}
\label{sec:conclusion}
We were able to show experimentally that MI displays a phase-dependent evolution, when stimulating the process with two seed fields, both of which are lying under the gain curve. Specifically, if the second modulation frequency matched a harmonic of the first modulation frequency, i.e. $\nu_{-2}=2\nu_{-1}$, an asymmetric evolution of the spectrum was observed. Thereby, this seeding concept provides a novel method to control MI-based frequency conversion and thus bears potential for applications, e.g. in optical parametric amplifiers the initial pump-seed phase difference could be used as a parameter to control the amplification. Or turning it around, a phase-dependent amplification might constitute a measure for phase changes of a weak seed signal. These applications are especially interesting as dichromatically seeded MI has been reported to exhibit low noise figures \cite{Andrekson2011}.
Furthermore, as MI dynamics dominate the initial stage of supercontinuum generation (SCG) in the long pulse regime, the presented findings should contribute to the understanding of SCG and anticipate a new method to influence SCG. Likewise, the presented results might be important for the understanding of other MI-based field evolutions such as the emergence of rogue waves \cite{Akhmediev2009c}.

\end{document}